\documentclass[12pt]{article}
\usepackage{authblk}
\usepackage[bookmarksnumbered, colorlinks, plainpages]{hyperref}
\usepackage{amsmath, amsthm, amscd, amsfonts, amssymb, graphicx, color, booktabs}
\usepackage{multirow}
\usepackage{subfigure}
\numberwithin{equation}{section}
\definecolor{email}{rgb}{0.00,0.00,0.84}
\begin{document}
\setcounter{page}{1}

\title{\large \bf 12th Workshop on the CKM Unitarity Triangle\\ Santiago de Compostela, 18-22 September 2023 \\ \vspace{0.3cm}
\LARGE Experimental status of $|V_{cd}|$ and $|V_{cs}|$}

\author[1]{Tengjiao Wang on behalf of the BESIII Collaboration}
\affil[1]{Nankai University, Tianjin, China}
%\dedicatory{This paper is dedicated to Professor ABCD}
\maketitle

\begin{abstract}
The Cabibbo-Kobayashi-Maskawa (CKM) matrix elements $|V_{cd}|$ and $|V_{cs}|$ are essential to test the unitary of CKM matrix. Before 2014, many results of $|V_{cd}|$ and $|V_{cs}|$ had been reported at BaBar, Belle, and CLEO experiments. The precisions of the most accurate results of $|V_{cd}|$ and $|V_{cs}|$ are 4.4\% and 3.3\%, respectively. After that, benefitting from larger data samples collected at BESIII, the precisions of $|V_{cd}|$ and $|V_{cs}|$ are improved to 1.8\% and 0.9\%, respectively.
\end{abstract} \maketitle
%---------------------------------------------------------------------------------------%
\section{Introduction}

%The Standard Model (SM), as the most optimal theoretical framework to date for describing the basic constituents of particles and their interactions, has undergone precise experimental tests.
The Cabibbo-Kobayashi-Maskawa~(CKM) matrix is an essential component of the Standard Model (SM) and can only be measured in experiment. It describes how quarks mix between their flavor eigenstates and weak interaction eigenstates when participating in weak interactions. The elements of the CKM matrix are expected to satisfy unitarity, ensuring that quarks in the fundamental particles exist in only three generations. Therefore, the precise measurements of CKM matrix elements are important to test the unitarity of the CKM matrix, thereby test the SM and search for possible new physics.

The (semi)-leptonic decays offer an important test-bed to investigate both the strong and weak interactions in the decays of charm quark. In the SM, the partial widths of the $D^+_{(s)}\to \ell^+\nu_\ell$ decay and the $D^+_{(s)}\to P\ell^+\nu_\ell$ decay  can be written as~\cite{fraction}
\begin{equation}
\Gamma_{D^+_{(s)}\to\ell^+\nu_\ell}=\frac{G_F^2}{8\pi}|V_{cd(s)}|^2
f^2_{D^+_{(s)}} m_{\ell^+}^2 m_{D^+_{(s)}} \left (1-\frac{m_{\ell^+}^2}{m_{D^+_{(s)}}^2} \right )^2
\end{equation}
and
\begin{equation}
\frac{\Gamma_{D^+_{(s)}\to P\ell^+\nu_\ell}}{dq}=\frac{G_F^2p^3}{24\pi^3}|V_{cd(s)}|^2
|f_+(q^2)|^2,
\end{equation}
respectively, where
$G_F$ is the Fermi coupling constant,
$m_\ell$ is the lepton mass,
$m_{D^+_{(s)}}$ is the $D^+_{(s)}$ mass,
$f_{D^+_{(s)}}$ is the $D^+_{(s)}$ decay constant, and
$|V_{cd(s)}|$ is the magnitude of the $c\to d(s)$
CKM matrix element. In recent years, BaBar, Belle, CLEO, and BESIII collaborations have reported lots of results of $|V_{cd(s)}|$ based on the (semi)-leptonic charm decays. This paper summarizes those experimental results of $|V_{cd}|$ and $|V_{cs}|$.
%---------------------------------------------------------------------------------------%
\section{Results before BESIII}
In 2008 and 2009, CLEO-c reported the results of $|V_{cd(s)}|$ by using $0.8(0.6)~\mathrm{fb}^{-1}$ of $e^+e^-$ collision data collected at the center-of-mass energy of $\sqrt{s}=3.774(4.170)$ GeV~\cite{CLEO-1,CLEO-2,CLEO-3,CLEO-4,CLEO-5}. In 2010,  using $0.52~\mathrm{ab}^{-1}$ of $e^+e^-$ collision data collected at the center-of-mass energy of $\sqrt{s}=10.6$ GeV, BarBar reported the results of $|V_{cs}|$~\cite{BarBar}. In 2013, Belle also reported the results of $|V_{cs}|$~\cite{BarBar}. The measured $|V_{cd(s)}|$ are summarized in Table~\ref{CLEO}.  Up to 2013, the most accurate results of $|V_{cd}|$ and $|V_{cs}|$ are obtained to be $|V_{cd}|=0.219\pm0.009\pm0.003$~\cite{CLEO-1} and $|V_{cs}|=1.107\pm0.019\pm0.028$~\cite{Belle}, respectively, and the corresponding precision of $|V_{cd(s)}|$ is 4.4(3.3)\%.
%Figure.~\ref{Belle} shows some typical distributions of $D_s^+\to\mu^+\nu_{\mu}$ and $D_s^+\to\tau^+\nu_{\tau}$ candidates at Belle collaboration~\cite{Belle}.

%\begin{figure}[htbp]
%   \centering
%   \includegraphics[width=2.5in]{figure/Belle1.png} 
%   \includegraphics[width=2.5in]{figure/Belle2.png} 
%   \includegraphics[width=2.5in]{figure/Belle3.png} 
%   \includegraphics[width=2.5in]{figure/Belle0.png} 
%   \caption{The upper left corner is the $M^2_{\rm miss}$ distribution of exclusively reconstructed $D_s^+\to\mu^+\nu_{\mu}$ decays within the inclusive $D_s^+$ sample superimposed fit results (solid blue line). The $E_{\rm ECL}$ distribution of exclusively reconstructed $D_s^+\to\tau^+(e^+)\nu_{\tau}$ (a), $D_s^+\to\tau^+(\mu^+)\nu_{\tau}$ (b), and $D_s^+\to\tau^+(\pi^+)\nu_{\tau}$ (c) decays within the inclusive $D_s^+$ sample with superimposed fit results~\cite{Belle}. }
%   \label{Belle}
%\end{figure}

\begin{table}[htbp]\centering
  \caption{Results of $|V_{cd(s)}|$ measured at CLEO-c, BarBar, and Belle.}
  \scalebox{0.8}{
  \begin{tabular}{l|c|c|c|c}
  \hline\hline
  Collaboration&Decay chain          &$N_{\rm sig}$      &$|V_{cd}|$                         &$\Delta$ (\%)\\
  \hline
\multirow{2}{*}{CLEO}  &$D^{+}\to\mu^{+}\nu_{\mu}$~\cite{CLEO-1}                &149.7 &$0.219\pm0.009\pm0.003$              &4.4\\
  &$D^{+(0)}\to\pi^{0(-)}e^{+}\nu_{e}$~\cite{CLEO-2}      &42(21) &$0.234\pm0.007\pm0.025$              &11.1\\
  \hline
 Collaboration& Decay chain                 &$N_{\rm sig}$      &$|V_{cs}|$                         &$\Delta$ (\%)\\
  \hline
\multirow{5}{*}{CLEO}   &$D^{+}_s\to\mu^{+}\nu_{\mu}$~\cite{CLEO-3}            &235.5  &$1.000\pm0.040\pm0.016$              &4.3\\
  &${D^{+}_s\to\tau^{+}\nu_{\tau}}^1$~\cite{CLEO-4}        &180.6  &$0.981\pm0.044\pm0.021$              &5.0\\
  &${D^{+}_s\to\tau^{+}\nu_{\tau}}^2$~\cite{CLEO-5}        &197.9  &$1.001\pm0.052\pm0.019$              &5.5\\
  &${D^{+}_s\to\tau^{+}\nu_{\tau}}^3$~\cite{CLEO-3}       &125.6  &$1.079\pm0.068\pm0.016$              &6.6\\
  &$D^{+(0)}\to K^-(\bar K^0) e^{+}\nu_{e}$~\cite{CLEO-2}  &27(54)  &$0.985\pm0.009\pm0.103$              &10.5\\
  \hline
\multirow{2}{*}{BarBar}   &$D^{+}_s\to\mu^{+}\nu_{\mu}$~\cite{BarBar}  &275 &$1.032\pm0.033\pm0.029$              &4.3\\
  &${D^{+}_s\to\tau^{+}\nu_{\tau}}^4$~\cite{BarBar}    &748 &$0.953\pm0.033\pm0.047$              &6.0\\
  \hline
\multirow{2}{*}{Belle}     &$D^{+}_s\to\mu^{+}\nu_{\mu}$~\cite{Belle}        &492 &$0.969\pm0.029\pm0.019$              &3.6\\
  &${D^{+}_s\to\tau^{+}\nu_{\tau}}^5$~\cite{Belle}    &2206 &$1.017\pm0.019\pm0.028$              &3.3\\
  \hline\hline
  \multicolumn{5}{l}{\bf{1 $\tau^{+}\to e^+\nu_e\bar{\nu}$; 2 $\tau^{+}\to\pi^+\pi^0\bar{\nu}$; 3 $\tau^{+}\to\pi^+\bar{\nu}$; 4 $\tau^{+}\to e(\mu)^+\nu_{e(\mu)}\bar{\nu}$;5 $\tau^{+}\to e^+\nu_{e}\bar{\nu}(\pi^+\bar{\nu},\mu^+\nu_{\mu}\bar{\nu})$}}\\
  \end{tabular}}
  \label{CLEO}
\end{table}

%---------------------------------------------------------------------------------------%
\section{Results after BESIII}
From 2004 to 2009, BEPC and BES have undergone major renovations and upgraded to BESIII and BEPCII, respectively. From 2010 to 2023, BESIII has collected large data samples at the center-of-mass energy of $\sqrt{s}=3.773$ GeV, 4.009 GeV, 4.13-4.23 GeV, and 4.6-4.7 GeV. 

Using 2.93 fb$^{-1}$ of data sample taken at 3.773 GeV, BESIII reported a series results of $|V_{cd}|$ and $|V_{cs}|$. In 2014, BESIII studied the $D^+\to\mu^+\nu_{\mu}$ decay~\cite{BES0}, with signal yield $409\pm21\pm2$, and the $|V_{cd}|$ of $|V_{cd}|=(0.2210\pm0.0058\pm0.0047)$. In 2019, BESIII firstly observed the $D^+\to\tau^+\nu_\tau$ signal~\cite{BES00}, with signal yield $137\pm27$, and the $|V_{cd}|$ of $|V_{cd}|=0.237\pm0.024\pm0.012\pm0.001$. From 2015 to 2020, BESIII studied some semi-leptonic decays of $D^0\to\pi^-e^+\nu_e$~\cite{BES1}, $D^+\to\pi^0e^+\nu_e$~\cite{BES2}, $D^+\to\eta e^+\nu_e$~\cite{BES4},  $D^+\to\eta\mu^+\nu_\mu$~\cite{BES5}, $D^0\to K^-e^+\nu_e$~\cite{BES1}, $D^+\to \bar K^0e^+\nu_e$~\cite{BES2}, $D^+\to K_L^0e^+\nu_e$~\cite{BES6}, and $D^0\to K^-\mu^+\nu_\mu$~\cite{BES7}. The measured $|V_{cd}|$ and $|V_{cs}|$ are summarized in Table.~\ref{BESIIIVcd1}. The most accurate result of $|V_{cd}|$ is given by Ref.~\cite{BES1}, corresponding to a precision of 1.8\%. And the most accurate results of $|V_{cs}|$ are given by Refs.~\cite{BES1,BES7}, corresponding to a precision of 0.9\%.

\begin{table}[htbp]\centering
  \caption{Results of $|V_{cd}|$ and $|V_{cs}|$ determined from semi-leptonic decays at BESIII, where the $|V_{cd}|$ are extracted by taking the form factors from the LQCD calculations as input. The values of $f^{\pi}_+(0)$ and $f^{K}_+(0)$ are cited from HPQCD2021, with precisions of 0.8\% and 0.6\%, respectively.}
  \begin{tabular}{l|c|c|c|c}
  \hline\hline
  Decay chain             &$f^{\pi}_+(0)|V_{cd}|$  &$f^{\pi}_+(0)$~\cite{Quote1}    &$|V_{cd}|$  &$\Delta$ (\%)\\
  \hline
  $D^0\to \pi^-e^+\nu_e$~\cite{BES1}         &$0.1435(18)(9)$         &\multirow{2}{*}{0.6300(51)}        &0.2278(34)(23)    &1.8\\
  $D^+\to \pi^0e^+\nu_e$~\cite{BES2}         &$0.1413(35)(12)$         &      &0.2243(58)(26)    &2.8\\
  \hline
 Decay Chain               &$f^{K}_+(0)|V_{cd}|$         &$f^{K}_+(0)$~\cite{Quote1}    &$|V_{cd}|$  &$\Delta$ (\%)\\
  \hline
  $D_s^+\to K^0e^+\nu_e$~\cite{BES3}         &$0.162(19)(3)$      &0.7452(31)        &0.217(26)(4)      &12.1\\
  \hline
  Decay Chain               &$f^{\eta}_+(0)|V_{cd}|$        &$f^{\eta}_+(0)$~\cite{Quote3}   &$|V_{cd}|$  &$\Delta$ (\%)\\
  \hline
  $D^+\to \eta e^+\nu_e$~\cite{BES4}          &$0.0815(45)(18)$        &\multirow{2}{*}{0.36(5)}           &0.2264(338)(318)    &20.5\\
  $D^+\to \eta\mu^+\nu_\mu$~\cite{BES5}        &$0.087(8)(2)$           &           &0.242(41)(34)           &21.8\\
  \hline
  Decay Chain               &$f^{K}_+(0)|V_{cs}|$    &$f^{K}_+(0)$~\cite{Quote1}    &$|V_{cs}|$  &$\Delta$ (\%)\\
  \hline
  $D^0\to K^-e^+\nu_e$~\cite{BES1}        &$0.717(03)(04)$         &\multirow{4}{*}{0.7452(31)}        &0.9622(57)(67)    &0.9\\
  $D^+\to \bar K^0e^+\nu_e$~\cite{BES2}         &$0.705(04)(11)$         &        &0.9461(67)(153)   &1.8\\
  $D^+\to K^0_Le^+\nu_e$~\cite{BES6}         &$0.728(06)(11)$         &        &0.9769(90)(153)   &1.8\\
  $D^0\to K^-\mu^+\nu_\mu$~\cite{BES7}        &$0.7148(38)(29)$        &        &0.9592(65)(56)    &0.9\\
  \hline
  Decay Chain               &$f^{\eta}_+(0)|V_{cs}|$        &$f^{\eta}_+(0)$~\cite{Quote2}   &$|V_{cs}|$  &$\Delta$ (\%)\\
  \hline
  $D_s^+\to \eta e^+\nu_e$~\cite{BES8}        &$0.446(10)(8)$           &0.495(5)           &0.9010(582)(569)      &9.0\\
  \hline
  Decay Chain              &$f^{\eta^{\prime}}_+(0)|V_{cs}|$        &$f^{\eta^{\prime}}_+(0)$~\cite{Quote2}   &$|V_{cs}|$  &$\Delta$ (\%)\\
  \hline
  $D_s^+\to \eta^{\prime} e^+\nu_e$~\cite{BES8}        &$0.477(100)(22)$            &$0.558^{+47}_{-45}$           &$0.8548(1920)^{+821}_{-794}$      &24.4\\
  \hline\hline
  %\multicolumn{5}{l}{\bf{(a) $f^{\pi}_+(0)$ and $f^{K}_+(0)$ are from PRD 107,094516 (2023).}}\\
 % \multicolumn{5}{l}{\bf{(b) $f^{\eta}_+(0)$ is from  Front. Phys. 14,64401 (2019).}}\\
   %\multicolumn{5}{l}{\bf{(a) $f^{K}_+(0)$ is from PRD 107,094516 (2023).}}\\
  %\multicolumn{5}{l}{\bf{(b) $f^{\eta}_+(0)$ is from JHEP 11,138 (2015).}}\\
  %\multicolumn{5}{l}{\bf{(c) $f^{\eta^{\prime}}_+(0)$ is from JHEP 11,138 (2015).}}\\
  \end{tabular}
  \label{BESIIIVcd1}
\end{table}

Using the data sample taken around 4.178 GeV, BESIII reported the results of $|V_{cd}|$ and $|V_{cs}|$ by using the $D_s^+$ (semi)-leptonic decays. In 2019, BESIII studied the decays of $D_s^+\to\mu^+\nu_{\mu}$~\cite{BES9}, $D_s^+\to \eta^{(\prime)}e^+\nu_e$~\cite{BES8}  and $D_s^+\to K^0e^+\nu_e$~\cite{BES3} with a 3.19 fb$^{-1}$ data sample taken at 4.178 GeV. In 2021, BESIII studied the leptonic decays of $D_s^+\to\mu^+\nu_\mu$~\cite{BES10}, $D_s^+\to\tau^+\nu_\tau(\tau^+\to\pi^+\bar\nu_\tau)$~\cite{BES10}, $D_s^+\to\tau^+\nu_\tau(\tau^+\to\rho^+\bar\nu_{\tau})$~\cite{BES11}, and $D_s^+\to\tau^+\nu_\tau(\tau^+\to e^+\nu_e\bar\nu_{\tau})$~\cite{BES12} by using a 6.32 fb$^{-1}$ data sample taken at 4.178-4.226 GeV. In 2023, BESIII reported three new results from the leptonic decays of $D_s^+\to\mu^+\nu_\mu$~\cite{BES13}, $D_s^+\to\tau^+\nu_\tau(\tau\to\pi\bar\nu_\tau)$~\cite{BES14}, $D_s^+\to\tau^+\nu_\tau(\tau^+\to\mu^+\bar \nu_\tau \nu_\mu)$~\cite{BES15}, $D_s^+\to\eta^{(\prime)} e^+\nu_e$~\cite{BES16}, and $D_s^+\to\eta^{(\prime)} \mu^+\nu_\mu$~\cite{BES17} using a 7.33 fb$^{-1}$ data sample taken at 4.128-4.226 GeV. The most accurate result of $|V_{cs}|$ in those $D_s^+$ leptonic decays is given by Ref.~\cite{BES13}, and the corresponding precision is 1.4\%.

%\begin{table}[htbp]\centering
%  \caption{Results of $|V_{cs}|$ from leptonic decay at BESIII}
%  \scalebox{0.9}{
%  \begin{tabular}{l|c|c|c|c|c}
%  \hline\hline
%  Reference       &$L/E_{\rm cm}$ (fb$^{-1}/$GeV)  &$\ell^+$  &$N_{\rm sig}$  &$|V_{cs}|$    &$\Delta$ (\%)\\
%  \hline
%  PRL122,071802  &3.19/4.178       &$\mu^{+}$      &$1136\pm33$  &$0.985(14)(14)$            &2.0\\
%  PRD104,052009  &6.32/4.178-4.226  &$\mu^{+}$      &$2198\pm55$  &$0.973(12)(15)(4)$         &2.0\\
%  arXiv:230.14585  &7.33/4.128-4.226  &$\mu^{+}$      &$2515\pm53$  &$0.968(10)(9)$         &1.4\\
%  \hline
%  PRD104,052009   &6.32/4.178-4.226 &${\tau^{+}}_a$ &$ 956\pm46$  &$0.972(23)(16)(4)$          &2.9\\
%  PRD104,032001  &6.32/4.178-4.226 &${\tau^{+}}_b$ &$1745\pm84$  &$0.981(44)(21)$             &3.1\\
%  PRL127,0171801 &6.32/4.178-4.226 &${\tau^{+}}_c$ &$4940\pm97$  &$0.978(9)(12)$              &{\color{red}1.5}\\
 %   arXiv:2303.12468 &7.33/4.128-4.226 &${\tau^{+}}_d$ &$2285\pm73$                &$0.987(16)(14)$         &2.2\\
 % arXiv:2303.12600 &7.33/4.128-4.226 &${\tau^{+}}_a$ &$2411\pm75$                &$0.993(15)(12)(4)$      &2.0\\
%  \hline\hline
%  \multicolumn{6}{l}{\bf{a $\tau^{+}\to\pi^+\bar{\nu}$; b $\tau^{+}\to\pi^+\pi^0\bar{\nu}$; c $\tau^{+}\to e^+\nu_e\bar{\nu}$;d $\tau^{+}\to \mu^+\nu_\mu\bar{\nu}$}}\\
%  \end{tabular}}
%      \label{BESIIIVcs2}
%\end{table}

%---------------------------------------------------------------------------------------%
\section{Summary}
\begin{figure}[htbp]
   \centering
   \subfigure[]{\includegraphics[width=0.45\textwidth]{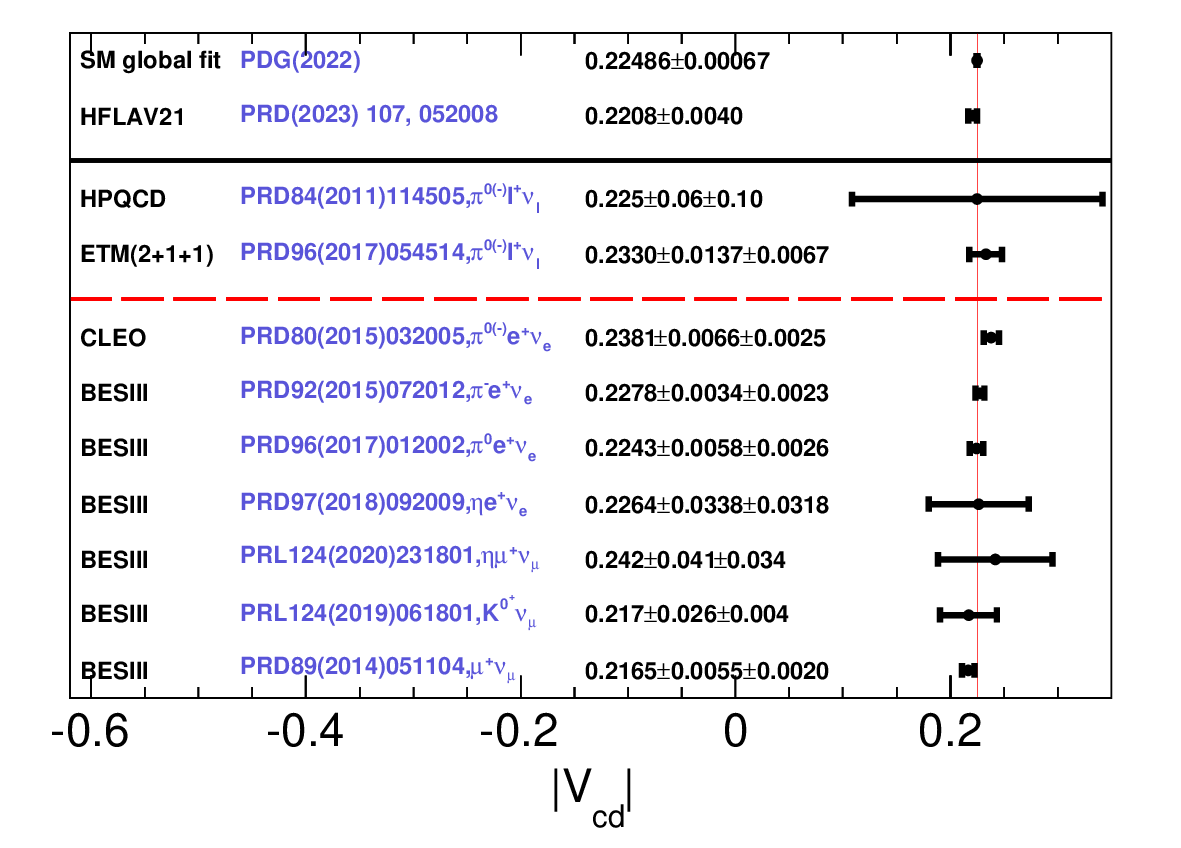}} \\
   \subfigure[]{\includegraphics[height=4.5cm]{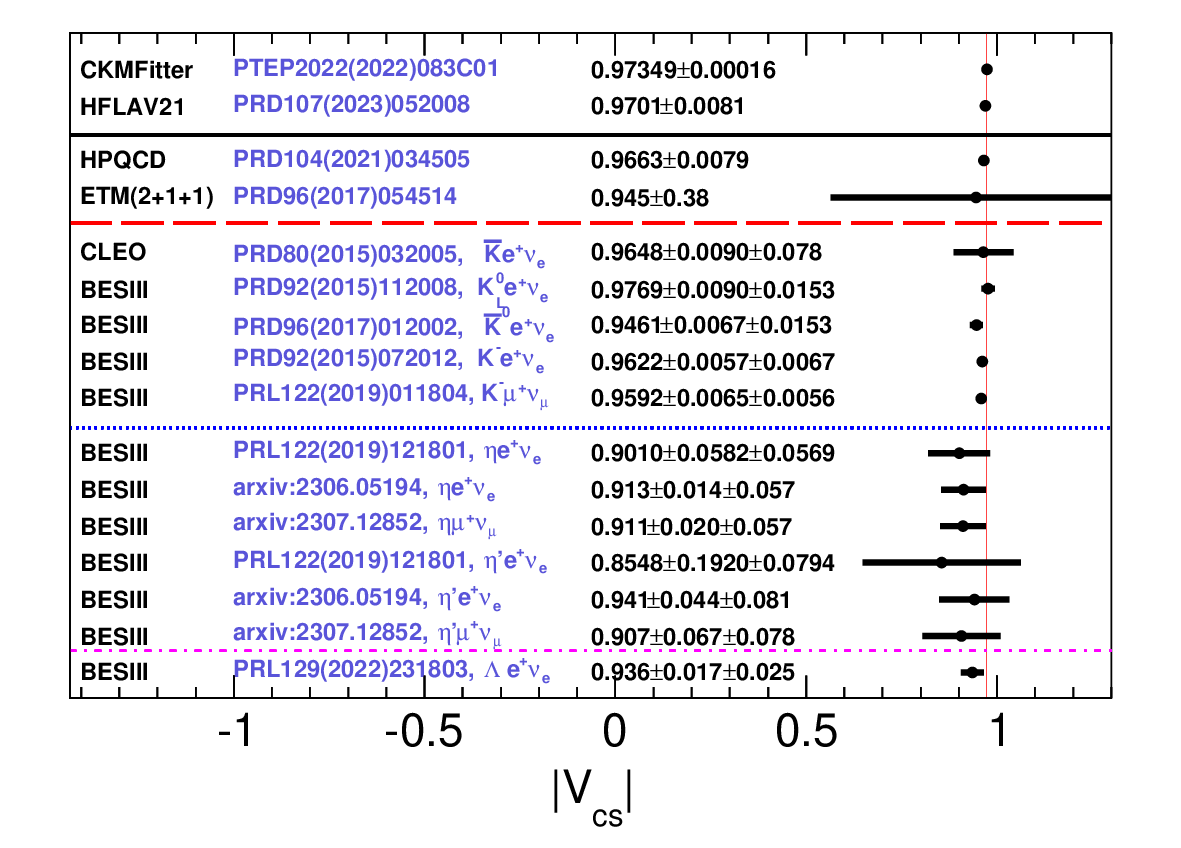} }
   \subfigure[]{\includegraphics[height=4.5cm]{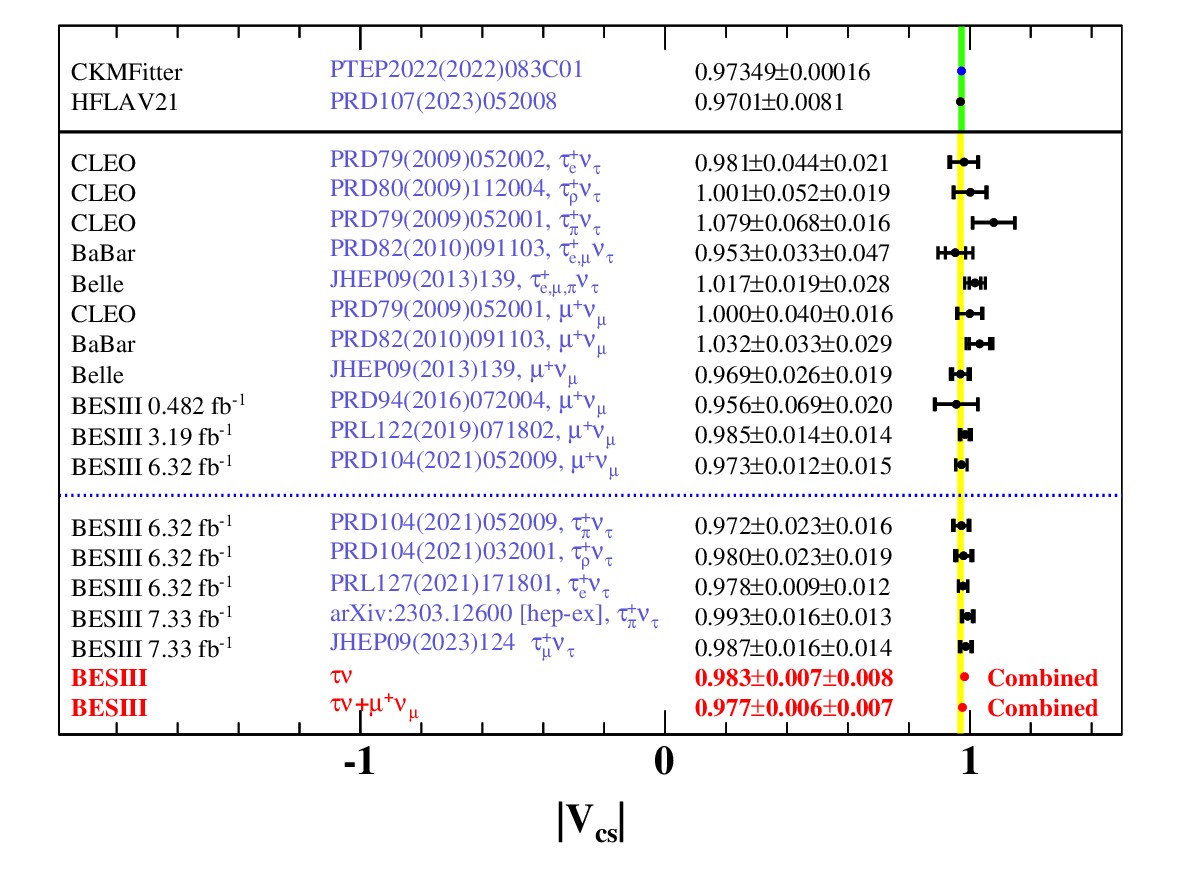}} 
   \caption{Comparisons of $|V_{cd}|$ and $|V_{cs}|$, where the $|V_{cs}|$ are recalculated using the same input values as Ref~\cite{BES13}, and the $|V_{cs}|$ are recalculated with the values of $G_F$, $m_{D^+}$, $m_{\tau}$, and $\tau_{\mu}$ from PDG2022.}
   \label{compare}
\end{figure}

From 2010 to 2023, BESIII systematically studied the leptonic and semi-leptonic decays of charmed mesons by using 2.93(7.33) fb$^{-1}$ data samples at the center-of-mass energy of $\sqrt{s}=3.773(4.128-4.226)$ GeV. The experimental measurement precisions of $|V_{cd}|$ and $|V_{cs}|$ are improved to 1.8\% and 0.9\%, respectively. The determined $|V_{cd(s)}|$ are consistent with the previous results. Figure~\ref{compare} shows the comparisons of the results from different experiments. From 2022 to 2024, BESIII has collected $\sim$17 fb$^{-1}$ of data sample at 3.773 GeV.  The precision of $|V_{cd}|$ and $|V_{cs}|$ will be further improved with the full 20 fb$^{-1}$ of data samples at 3.773 GeV at BESIII.

%---------------------------------------------------------------------------------------%
\bibliographystyle{amsplain}

\end{document}